\documentclass[aps,prl,floatfix,twocolumn,nofootinbib]{revtex4}
\usepackage{dcolumn}
\usepackage{bm}
\usepackage[dvips]{epsfig}
\usepackage{graphicx} \usepackage{amsmath} \usepackage{amssymb}

\def\bea{\begin{eqnarray}}
\def\eea{\end{eqnarray}}

\newcommand{\comment}[1]{}
\newcommand{\BEQ}{\begin{equation}}
\newcommand{\EEQ}{\end{equation}}
\newcommand{\BEA}{\begin{eqnarray}}
\newcommand{\EEA}{\end{eqnarray}}
\renewcommand{\d}{{\rm d}}

\begin{document}

\title{Bargaining with entropy and energy
}
\author{S.G. Babajanyan$^{1,2}$, K.H. Cheong$^2$, and A.E. Allahverdyan$^1$}

\affiliation{$^1$Alikhanyan National Laboratory (Yerevan Physics
  Institute), Alikhanian Brothers Street 2, Yerevan 375036, Armenia}
\affiliation{$^2$Singapore University of Technology and Design, 8 Somapah Road, Singapore}

\begin{abstract} 

Statistical mechanics is based on interplay between energy minimization
and entropy maximization. Here we formalize this interplay via axioms of
cooperative game theory (Nash bargaining) and apply it out of
equilibrium. These axioms capture basic notions related to joint
maximization of entropy and minus energy, formally represented by
utilities of two different players. We predict thermalization of a
non-equilibrium statistical system employing the axiom of affine
covariance|related to the freedom of changing initial points and
dimensions for entropy and energy|together with the contraction
invariance of the entropy-energy diagram. Whenever the initial
non-equilibrium state is active, this mechanism allows thermalization to
negative temperatures. Demanding a symmetry between players fixes the
final state to a specific positive-temperature (equilibrium) state.  The
approach solves an important open problem in the maximum entropy
inference principle, {\it viz.} generalizes it to the case when the
constraint is not known precisely. 


\end{abstract}


\maketitle

Interplay between entropy and energy is fundamental for
equilibrium statistical mechanics
\cite{ingo,callen,landau,lindblad,mahler}. The interplay is based on the
fact that the equilibrium (positive-temperature) Gibbs distribution can
be obtained via maximizing entropy for a fixed energy, or via minimizing
energy for a fixed entropy \cite{ingo,callen,landau,lindblad,mahler}.
The entropy maximization reflects the tendency of an isolated system
towards maximal disorder. The energy minimization relates to finding
a more stable (passive) state \cite{lindblad}. 

Here we show that the competition between entropy and energy can be
formalized via axioms of cooperative game theory (bargaining)
\cite{luce,roth_book,roth1} and applied out of equilibrium. Now entropy $S$
and minus energy $U=-E$ are payoffs of two players that tend to
maximize them ``simultaneously"; see Table I of \cite{sup} 
for a game theory|statistical physics dictionary. While non-cooperative game theory
focuses on rational actions to be chosen given payoffs only, the
bargaining provides an axiomatic description of interaction between the
players that should reach a compromise \cite{nash,luce,roth_book}. Hence
bargaining is suitable for formalizing interaction mechanisms to be
applied in physics \cite{foo1}. We apply it when known thermodynamic
principles do not suffice for predicting system's behavior. Given a
non-equilibrium initial state with entropy larger than $k_{\rm B}\ln 2$,
and using certain plausible axioms on the relaxation process, we can
show that the final state is an equilibrium one, with the sign of
temperature depending on the initial state. The final state is fixed to
a specific positive-temperature state, if the symmetry between players is
assumed. Thus we derive thermalization via game theory. As an
application of our results we resolve a major open problem in the maximum entropy 
inference method \cite{jaynes,maxent_action} generalizing it to those cases, where the
constraint is not known precisely.

\begin{figure}[ht]
\includegraphics[width=8cm]{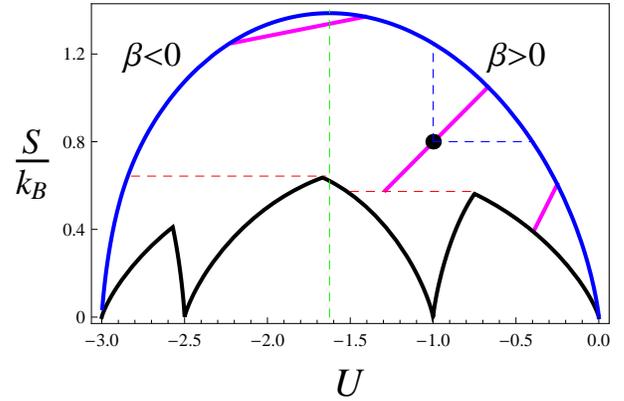}
\caption{A typical example of entropy-energy diagram.
Entropy is $S$ and the minus energy $U=-E$ for 4-level system with energies:
$\varepsilon_1=0$, $\varepsilon_2=1$, $\varepsilon_3=2.5$, and
$\varepsilon_4=3$. Maximal (minimal) entropy curves are denoted by blue
(black). All physically acceptable values of entropy and energy are
inside of the domain bounded by blue and black curves. States below 
red dashed lines (both are lower than $\ln 2$) are excluded by Axiom 2. 
Green dashed line shows $U_{\rm av}=-\frac{1}{4}\sum_{k=1}^4\varepsilon_k$; it separates $\beta>0$ from $\beta<0$; cf.~(\ref{gop}). 
Black point denotes a possible initial state. States inside dashed blue lines hold axiom 4.
Magenta lines denote initial states that produce the same final state (\ref{bolo}). 
}
\label{fig1}
\end{figure} 

{\it Entropy-energy diagram}. We study a classical system with discrete
states $i=1,..n$ and respective energies $\{\varepsilon_i\}_{i=1}^n$. A
statistical (generally non-equilibrium) state of the system is given by probabilities  
\BEA
\label{simpo}
\{p_i\geq 0\}_{i=1}^n, \qquad {\sum}_{i=1}^n p_i=1.
\EEA
The entropy and minus average energy for such a state read, respectively
\cite{ingo,callen,landau,lindblad,mahler}:
\BEA
\label{bugi}
S[p]=-k_{\rm \,B}{\sum}_{i=1}^n p_i\ln p_i, \qquad
U[p]=-{\sum}_{i=1}^n \varepsilon_i p_i,
\EEA
where $k_{\rm \,B}$ is Boltzmann's constant. 
The Gibbsian
equilibrium states are obtained by maximizing $S[p]$ over (\ref{simpo}) under a fixed $U=U[p]$
\cite{ingo,callen,landau,lindblad,mahler}:
\BEA
\label{gi}
\pi_i=e^{-\beta \varepsilon_i}/Z, \quad Z={\sum}_{i=1}^n e^{-\beta \varepsilon_i},\quad \beta=1/(k_{\rm \,B} T),
\EEA
where the inverse temperature $\beta$ is uniquely
determined from $U[p]=U$. The same result|but restricted to the
positive-temperature branch $\beta>0$|is obtained upon maximizing $U[p]$
under fixed $S$ \cite{lindblad,mahler}. This is why we frequently employ
the minus energy together with entropy: both are maximized in
equilibrium.

Fig.~\ref{fig1} shows a typical entropy-energy diagram 
on the $(U,S)$ plane. The maximum entropy curve $S(U)$
holds
\BEA
S(U_{\rm av})=\ln n, \quad U_{\rm av}\equiv-\frac{1}{n}{\sum}_{k=1}^n 
\varepsilon_k.
\label{gop}
\EEA
For $U>U_{\rm av}$ ($U<U_{\rm av}$) $S(U)$ refers to $\beta>0$ ($\beta<0$). 
No probabilistic states
are possible below the minimum entropy curve $S_{\rm min}(U)={\rm
min}_{p, \,U[p]=U}(S[p])$. The maximum entropy curve is smooth and bounds a convex
domain due to concavity of $S[p]$ [$0<\epsilon<1$]
\cite{lindblad,mahler}: $S[\epsilon p+(1-\epsilon)q]\geq \epsilon S[
p]+(1-\epsilon)S[q]$. Now $S_{\rm min}(U)$ is an irregular curve,
because minima of a concave function $S[p]$ are reached for vertices of
the allowed probability domain that combines (\ref{simpo}) 
with constraint $U[p]=U$. Hence only two probabilities are
non-zero; see \S 1 of \cite{sup} for details. We have
[cf.~Fig.~\ref{fig1}]:
\BEA
\label{tars}
S_{\rm min}(U)\leq k_{\rm \,B} \ln 2, \qquad S_{\rm min}(-\varepsilon_i)=0.
\EEA

\comment{

Generically, the bound is not reached, though $S_{\rm min}(U)$ can be
close to it for certain values of $U$, as Fig.~\ref{fig1} shows.

Below we shall restrict ourselves with probabilistic states whose
entropy is larger than $k_{\rm \,B}\ln 2$. Those sufficiently
macroscopic states are in the application domain of game-theoretic
axioms we explore below.  For concrete values of
$\{\varepsilon_i\}_{i=1}^n$ this restriction can be loosened, e.g. for
the situation given in Fig.~\ref{fig1} we can consider all probabilistic
states above the dashed red lines; both these lines have entropies lower
than $k_{\rm \,B}\ln 2$. }

{\it Statement of the problem.} We emphasize that all above features of
the entropy-energy diagram hold for arbitrary large, but finite values
of $n$. Let the statistical system be found initially at some point of
the entropy-energy diagram. We want to predict the long-time state of
this system, knowing that its entropy and minus energy tend not to
decrease. An example of this situation is when a thermally isolated
statistical system is subject to external fields that extract energy.
(Recall that any process is thermally isolated if the environment is
included into the system.) Now in two extreme cases thermodynamics can
determine the long-time state \cite{ingo,callen,landau}: firstly, if the
work-extraction process is very slow, then the entropy is conserved and
work-extraction entails decreasing energy.  Consequently, through the
extraction of as much work as possible, the system will finally reach
equilibrium along the constant entropy \cite{ingo,callen,landau}.
Secondly, when no work-extraction is present and the system is
completely isolated, its entropy will increase till it finally reaches
equilibrium along the constant energy path.

\comment{
If the initial minus-energy is lower than the average minus-energy
$U<-\frac{1}{n}\sum_{k=1}^n \varepsilon_k$, then this equilibrium is
reached at a negative temperature. Otherwise, for
$U>-\frac{1}{n}\sum_{k=1}^n \varepsilon_k$, the system will reach a
positive-temperature equilibrium state. }

Now what if both processes occur simultaneously, i.e. when both
entropy and minus energy increase, can one still show that the system
will reach a thermal equilibrium?  If yes, can one bound its
temperature? 

The standard thermodynamics cannot answer these two questions due to
insufficient information. (E.g. it can predict the final state (\ref{gi}) 
if we know that the system is attached to a thermal bath at inverse 
temperature $\beta$; but we do not make such an assumption.) 
The questions can be answered within more
detailed, non-equilibrium statistical mechanical theories \cite{mahler,
lindblad}. But such theories make a number of dynamical assumptions,
e.g. they assume that internal constituents of the system move according
to quantum Hamiltonian dynamics during the whole system's evolution
\cite{lindblad,mahler}. Or they assume that the systems is of
hydrodynamic type with smooth density, velocity and pressure fields
\cite{ingo}. The direct validity of such assumptions is difficult to
address, hence a specific axiomatic approach is required
\cite{foo2}. 

Here we address the above question via axioms of bargaining games
\cite{nash,luce,roth_book}; see Table I of \cite{sup} for a detailed
comparison between bargaining theory and statistical physics. Given the
initial state $(U_{\rm i}, S_{\rm i})$, we look for the final state
$(U_{\rm f}, S_{\rm f}) =(U[p_{\rm\, f}], S[p_{\rm \,f}])$. Axioms below
will determine this final state on the entropy-energy diagram. 

{\it Axiom 1}: Once both $S$ and $U$ tend to increase, then 
at least one of them should increase to some extent:
\BEA
U_{\rm i}\leq U_{\rm f}, \qquad S_{\rm i}\leq S_{\rm f},
\label{sumo}
\EEA
where at least one inequality is strict. 
In game theory (\ref{sumo}) relates to individual rationality of players 
\cite{roth_book,roth1}. 

{\it Axiom 2: The choice of initial conditions.} We shall assume a
non-equilibrium initial (probabilistic) state, with entropy $S_{\rm
i}>k_{\rm B}\ln 2$. This is a class of sufficiently macroscopic states
for our purposes.  In concrete cases this condition can be made weaker;
e.g. for the case of Fig.~\ref{fig1} we can allow all initial states
above dotted red lines.  Now Axioms 1 and 2 ensure that the domain of
allowed final states on the entropy-energy diagram is a convex set. 

{\it Axiom 3: Affine-covariance.} If the entropy-energy diagram $(U,S)$ (including 
$(U_{\rm i}, S_{\rm i})$) is transformed as
\BEA
\label{a1}
( U,S )\to ( a^{-1} U +d, b^{-1} S +c ),\qquad a>0,\, b>0,
\EEA
where $c$ and $d$ are arbitrary, then the final state is transformed via
the same rule (\ref{a1})  \cite{luce}. The freedom to translate energy and entropy
by an arbitrary amount is well-known in physics; hence the factors $c$
and $d$ in (\ref{a1}). In (\ref{a1}), $a$ and $b$ account for
the fact of different dimensions for $S[p]$ and $U[p]$ [cf.~(\ref{bugi})], and the
possibility of changing those dimensions without changing physics. We shall apply
(\ref{a1}) also for dimensionless $a$ and $b$. 

\comment{
In game theory
(\ref{a1}) reflects the freedom related to the expected utility
theory \cite{luce}. 
\BEA
\label{a11}
(U_{\rm f}, S_{\rm f} )\to ( a^{-1} U_{\rm f} +c, b^{-1} S_{\rm f}+d).
\EEA
}

{\it Axiom 4: Contraction invariance (Independence of irrelevant
alternatives)} \cite{luce}. Let ${\cal D}'$ is a subset of the original
entropy-energy phase-diagram ${\cal D}$, and ${\cal D}'$ contains both
$(S_{\rm i}, U_{\rm i})$ and $(S_{\rm f}, U_{\rm f})$. If now the set of
allowed final states is restricted (contracted) from ${\cal D}$ to ${\cal
D}'$, then it still holds that $(U_{\rm i}, S_{\rm i},)\to (U_{\rm f},
S_{\rm f})$. 

Axiom 4 tells about any subset ${\cal D}'$, but below we shall need it
only for full-measure, well-behaved subsets that are similar to ${\cal
D}$. The restriction to those subsets can be realized via suitable
external fields. 

Contraction invariance plays an important role in decision theory
\cite{nash,luce,roth_book}. The intuition behind this axiom is {\it
physical}: it assumes that the actual evolution $(U_{\rm i}, S_{\rm
i})\to (U_{\rm f}, S_{\rm f})$ amounts to selecting the ``best" state
via binary comparisons of diagram points.  Hence
restricting the set of alternatives|provided that the ``best" states is
still allowed|cannot change the ``best".

\begin{figure}[ht]
\includegraphics[width=6cm]{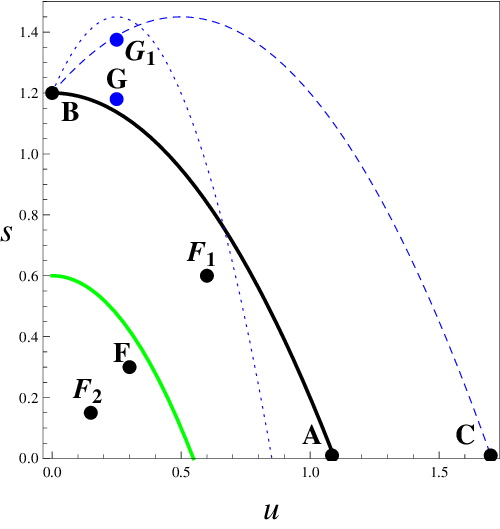}
\caption{The set of states in coordinates (\ref{ojgur}) for
$U_{\rm i}>U_{\rm av}$ (black and green curves) and 
$U_{\rm i}<U_{\rm av}$ (dotted and dashed blue curves). 
In both cases the initial state is shifted to $(0,0)$ via (\ref{a1}). 
${\bf F}$ and ${\bf G}$ are tentative final states for (resp.) first and second
scenario. 
}
\label{fig3}
\end{figure} 

{\it Thermalization.} We have two possibilities
for the initial state $(U_{\rm i}, S_{\rm i})$ [see (\ref{gop}) and Fig.~\ref{fig3}, \ref{fig1}]:
$U_{\rm i}\geq U_{\rm av}$ or $U_{\rm i}<U_{\rm av}$. Both scenarios are studied in coordinates
\BEA
\label{ojgur}
u=U-U_{\rm i}, \qquad s=S-S_{\rm i}, 
\EEA
which is done by (\ref{a1}) with $a=b=1$; see Fig.~\ref{fig3}.

For $U_{\rm i}\geq U_{\rm av}$, all points on and below the maximum
entropy curve ${\bf BA}$ [apart of $(0,0)$] are allowed as possible
final states; see (\ref{sumo}) and Fig.~\ref{fig3}. Assume that the
final state ${\bf F}\equiv(\bar{u}, \bar{s})\not \in {\bf BA}$;
cf.~Fig.~\ref{fig3}. Then there is a point ${\bf F_1}=(a_1\bar{u},
b_1\bar{s})$, with $a_1\geq 1$ and $b_1\geq 1$ (at least one of these
inequalities is strict). We now apply (\ref{a1}) with $a=a_1$, $b=b_1$,
and $c=d=0$ to all point of the diagram. This transforms ${\bf F}_1\to
{\bf F}$, and changes the original domain ${\cal D}={\bf OBA}$
[$(0,0)={\bf O}$] to smaller domain ${\cal D}'\subset {\cal D}$; see
Fig.~\ref{fig3}, where ${\cal D}'$ is below the green line. Now ${\bf
F}\in{\cal D}'$, due to ${\bf F}_1\to {\bf F}$. But we can regard ${\cal
D}'$ as just a subset of $\in{\cal D}$ and apply to it Axiom 4.  We now
have 2 contradicting facts: following Axiom 4, ${\bf F}$ should not
change when going to a subset. But it should change, ${\bf F}\to {\bf
F}_2$ according to (\ref{a1}), i.e. Axiom 3.  The contradiction is
avoided only if ${\bf F}$ is located on the maximum entropy curve ${\bf
BA}$ \cite{roth1}. 

For $U_{\rm i}<U_{\rm av}$, we can apply (\ref{a1}) only with $a=1$ and
$b>1$ ($c=d=0$). Otherwise, ${\cal D}'\not\subset{\cal D}$|see
Fig.~\ref{fig3} for an example|and then Axiom 4 does not apply.
Hence given a tentative final state ${\bf G}=(\bar{u},\bar{s})$ we can
reach the above contradiction only if there is a point ${\bf
G}_1=(\bar{u},b\bar{s})$; see Fig.~\ref{fig3}. I.e. the set of possible
final states coincides with curve ${\bf BC}$ in Fig.~\ref{fig3}. Note 
that for this conclusion we should slightly modify Axiom 1:
$U_{\rm i}> U_{\rm f}$; cf.~(\ref{sumo}).

Hence Axioms $1-4$ imply thermalization: the final state is on the
maximum entropy curve. Negative-temperature states are allowed by this
derivation for $U_{\rm i}<U_{\rm av}$. Such initial states are active,
i.e.  $(p_i-p_j)(\varepsilon_i-\varepsilon_j)>0$ at least for one pair
$(i,j)$. 

More information on the final state is contained in

{\it Axiom 5: Symmetry}: We made $U$ and $S$ dimensionless via
(\ref{a1}). If the domain of allowed final states (\ref{sumo}) is
symmetric|i.e. it contains a point $(U,S)$ if and only if it contains
$(S,U)$|and so is the initial state $(U_{\rm i}= S_{\rm i})$, then the
final state is also symmetric $(S_{\rm f}, S_{\rm f})$,
provided that there are no reasons to
regard the players asymmetrically. 

Nash \cite{nash} and his followers \cite{roth_book} argued 
that the only final state satisfying axioms $1-5$ is
\BEA
\label{bolo}
(U_{\rm N}, S_{\rm N})={\rm argmax}_{(U,S)}\left[(U-U_{\rm i})(S-S_{\rm i})   \right],
\EEA
where the maximum is reached on the maximum entropy curve restricted by
(\ref{sumo}). Since this curve is concave, the argmax in (\ref{bolo}) is
unique; see \S 2.1 of \cite{sup}. Eq.~(\ref{bolo}) shows that
$(U_{\rm N}, S_{\rm N})$ refer to a $\beta_N>0$; cf.~(\ref{gi}). 

\comment{For if it leads to two equivalent outcomes, we easily reach a
contradiction by showing that the middle point of the outcomes has a
higher value of $(U-U_{\rm i})(S-S_{\rm i})$. }

\begin{figure}[ht]
\includegraphics[width=6cm]{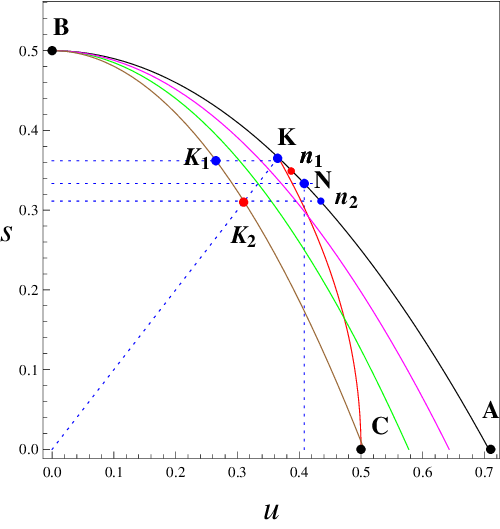}
\caption{ Entropy-energy diagram in coordinates (\ref{ojgur}): $s(u)$ (black),
$s(a_0u)$ (green), $s(a_1u)$ (magenta) and $s(a_2u)$ (brown). 
}
\label{fig5}
\end{figure}

However, Refs.~\cite{nash,roth_book} derive (\ref{bolo}) by making an
additional assumption, {\it viz.} the domain restricted by (\ref{sumo})
can enlarged into a larger domain; see \S 2.2 of \cite{sup} for details. We
cannot employ this assumption, since it is completely unphysical.  We
shall derive (\ref{bolo}) using axioms $1-5$, but without the
assumption.  Fig.~\ref{fig5} shows a typical example of the maximal
entropy curve $s(u)$|denoted by {\bf BA} in Fig.~\ref{fig5}|in
coordinates (\ref{ojgur}) with ${\bf O}\equiv (0,0)$.  The domain of
states {\bf BAO} allowed by Axiom 1 is not symmetric in the sense of
Axiom 5. But it has the largest symmetric subset ${\bf OBKC\subset
OBA}$, where ${\bf K}$ is the solution of $s(\hat u)=\hat u$, and ${\bf
KC}$ is the inverse function $s^{-1}(u)$ of $s(u)$; see Fig.~\ref{fig5}.
For ${\bf OBKC}$ Axiom 5 + thermalization lead to ${\bf K}$ as the final state.
Hence for the original domain ${\bf BAO}$ the final state is located on
the line ${\bf KA}$ (possibly including {\bf K}); see Axiom 4 and the result that the final state is
located on {\bf BA}. In coordinates (\ref{ojgur}) the state (\ref{bolo})
is given as $(u_N,s(u_N))$ with $u_N={\rm argmax}_u[us(u)]$.  This is
the point {\bf N} on Fig.~\ref{fig5}. Now ${\bf N}\in{\bf KA}$, as seen
from working out concave function $s(u)$ for $u\simeq\hat u$:
\BEA
\label{tot1}
 us(u)-\hat us(\hat u)=\hat u(u-\hat u)[\,1+s'(\hat u)\,],\\
 s(u)-s^{-1}(u)=\left[s'(\hat u)^2-1\,\right]\,(u-\hat u)\,\left/\,s'(\hat u)\right.,
\label{tot2}
\EEA
where $s'(u)=\d s/\d u$ and
factors ${\cal O}[(u-\hat u)^2]$ were neglected. Now for $-1<s'(u)<0$ we have the
situation shown on Fig.~\ref{fig5}, where $u_N>\hat u$ and $ s(u)>s^{-1}(u)$ for
$u>\hat u$. For $-1>s'(u)$ we have the analogue of Fig.~\ref{fig5}, where $u_N<\hat u$,
and the solution is located on ${\bf BK}$. For $1>s'(\hat u)>0$, we always get $u_N>\hat u$.  

The same restricting of the
final state will be done after transformation (\ref{a1}) with $b=1$ and
$c=d=0$, where $s(u)\to s(au)$. We choose $a=a_0$ such that
\BEA
{\rm argmax}_u[\, u\,s(a_0u)\,]=\hat u_0, \quad \hat u_0=s(a_0 \hat u_0),
\EEA
i.e. the transformed Nash solution (\ref{bolo}) equals $\hat u_0$. Consider
(\ref{a1}) under two other values of $a$: $a_2>a_0>a_1$; see Fig.~\ref{fig5}. 
When applying (\ref{a1}) with
$a=a_2$, the transformed Nash solution lies on ${\bf BK_2}$. 
On that curve, the final state lies between ${\bf K_1}$ and ${\bf K_2}$.
Going back to the original curve ${\bf KA}$, we restrict the final
state to lay between ${\bf K}$ and ${\bf n}_2$ on ${\bf KA}$. 
Applying (\ref{a1}) with $a=a_1$, we further restrict the final state to lay
between ${\bf n}_1$ and ${\bf n}_2$; see Fig.~\ref{fig5}. For
$a_1\to a_0\leftarrow a_2$ we get ${\bf n_1}\to{\bf N}\leftarrow {\bf
n_2}$, i.e. the final state coincides with (\ref{bolo}). 

{\it Retrodicting from an equilibrium state (without conservation
laws).} Given an equilibrium state $(U(\beta),S(\beta))$ with $\beta>0$
we can identify it with the final state (\ref{bolo}), and ask which
initial states give rise to it. Such a question is possible to ask
within standard thermodynamics only if the conservation law of entropy
or energy is there.  Note that $(U_{\rm N}, S_{\rm N})$ in
(\ref{bolo}) is determined from $\frac{\d}{\d U}\left[(U-U_{\rm
i})(S(U)-S_{\rm i}) \right]=0$. This leads to a line $S-S(\beta)=\beta
(U- U(\beta))$ on the entropy-energy diagram; see magenta lines on
Fig.~\ref{fig1}. They start from $(U(\beta),S(\beta))$ and end either at
a boundary of the diagram, or at a point, where the
convexity of the domain is lost. 

{\it The maximum entropy method} grew out of statistical physics and is
widely used in probabilistic inference \cite{jaynes,maxent_action,jaynes-stand}. To
recall it: probabilities (\ref{simpo}) are not known, but the average
value $U=U[p]\equiv-\sum_{k=1}^np_k\varepsilon_k$ of a random quantity
$\{-\varepsilon_k\}_{k=1}^n$ is known. Then the most unbiased (least
informative) probabilities that correspond to this prior information are
derived by maximizing the entropy $S[p]$ under $U=U[p]$:
${\rm max}_pS[p]=S( U)$ \cite{maxent_action,jaynes-stand,uffink}.
Applications of the method meet the following problem: if
$\{p_k\}_{k=1}^n$ are not known, and $U$ is learned from experiments,
then it cannot be known precisely \cite{jaynes-stand,uffink}; see
\S 3 of \cite{sup} for a review of the method and
the open problem. Thus we ask how is the method applied if there is
weaker prior information, i.e. the average $U$ is not known precisely,
but we only know that it belongs to the interval $U\in[U_1,U_2]$? For technical clarity we add another 
constraint [see (\ref{gop}) and \S 3.4 of \cite{sup} for extensions]:
\BEA
\label{bobo}
U\in [U_1,U_2],\qquad U_{\rm av}<U_1<U_2.
\EEA
Recall that upon (\ref{bobo}) the same result is obtained by maximizing $S[p]$
for $U=U[p]$, or by maximizing $U[p]$ for $S=S[p]$. 
Note that the uncertainty (\ref{bobo}) translates into the uncertainty $S\in [S_1,S_2]$
($S_1\equiv S(U_1)$, $S_2\equiv S(U_2)$) for the maximum entropy. 
For applying the bargaining axioms|where $U[p]$ and $S[p]$ are two utilities that tend to maximize simultaneously|we 
should define the domain of allowed states $\Omega$ on the $(U,S)$-diagram. Now
$\Omega$ is defined by joining two uncertainty intervals:
\BEA
\Omega=\{(\bar{U}, \bar{S})\, |\,\bar{U}\in [U_1,U_2],\, \bar{S}\in [S_1,S_2] \,\}.
\EEA
$\Omega$ has the form required by Axiom 1 with the initial point 
$(U_{\rm i}, S_{\rm i})= (U_1,S_2)$; cf.~(\ref{sumo}). Now Axioms 1$-$5 apply; see \S 3.3 of \cite{sup} for details.
Thus we deduce the bargainig solution for the uncertainty given by (\ref{bobo}) [cf.~(\ref{bolo})]:
\BEA
\label{kogan}
{\rm argmax}_{(U,S)}\left[  (U-U_{1})(S-S_2) \right]. 
\EEA
In particular, Axiom 5 is natural here due to the duality
between maximizing $U[p]$ and $S[p]$. Extensions of (\ref{kogan}) to
other [than (\ref{bobo})] uncertainty intervals $[U_1,U_2]$ are worked
in \S 3.4 of \cite{sup}. Elsewhere we shall explore applications of this
generalized maximum-entropy method.

\acknowledgements

We are grateful to E. Pogossian for motivating us to look at this
subject. We thank K. Hovhannisyan for useful discussions. AEA and SGB
were supported by SCS of Armenia, grants No. 18RF-002 and No. 18T-1C090.



\section{Supplementary Material}

\subsection{1. Calculation of the minimum entropy for a fixed average energy} 

Here we show how to mimimize entropy
\BEA
\label{entro}
S_{\rm min}(E)={\rm min}_p S[p],
\quad S[p]=-k_{\rm \,B}{\sum}_{i=1}^n p_i\ln p_i, 
\EEA
over probabilities
\BEA
\label{simplex}
\{p_i\geq 0\}_{i=1}^n, \qquad {\sum}_{i=1}^n p_i=1,
\EEA
for a fixed average energy
\BEA
\label{eno}
E={\sum}_{i=1}^n \varepsilon_i p_i.
\EEA
Energy levels $\{\varepsilon_i\geq 0\}_{i=1}^n$ are given. 

Since $S[p]$
is concave, its minimum is reached for vertices of the allowed
probability domain. This domain is defined by the intersection of
(\ref{simplex}) with probabilities that support constraint (\ref{eno}).
Put differently, as many probabilities nullify for the minimum of
$S[p]$, as allowed by (\ref{eno}). Hence at best only two probabilities
are non-zero.  

We now order different energies as 
\BEA
\varepsilon_1<\varepsilon_2<\varepsilon_3..., 
\EEA
and
define entropies $s_{ij}(E)$, where only states $i$
and $j$ with $i<j$ have non-zero probabilities:
\begin{gather}
\label{21}
s_{ij}(E)=-\frac{E-\varepsilon_i}{\varepsilon_j-\varepsilon_i}\ln \frac{E-\varepsilon_i}{\varepsilon_j-\varepsilon_i}
-\frac{\varepsilon_j-E}{\varepsilon_j-\varepsilon_i}\ln \frac{\varepsilon_j-E}{\varepsilon_j-\varepsilon_i},\\
\varepsilon_i\leq E\leq \varepsilon_j.
\end{gather}
The minimum entropy $S_{\rm min}(E)$ under (\ref{eno}) is found by looking|for a fixed $E$|at the minimum over all 
$s_{ij}(E)$ whose argument supports that value of $E$. E.g. $S_{\rm min}(E)$
reads from (\ref{21}) for $n=3$ (3 different energies):
\begin{gather}
S_{\rm min}(E)={\rm min}[\,     s_{13}(E),\,      \nonumber\\
\theta(\varepsilon_2-E)s_{12}(E) +\theta(E-\varepsilon_2)s_{23}(E)\,],
\label{werwolf}
\end{gather}
where $\theta(x)$ is the step-function ($\theta(x>0)=1$ and $\theta(x<0)=0$), and where
we assume $\varepsilon_1\leq E\leq \varepsilon_3$. Likewise, for $n=4$:
\begin{gather}
S_{\rm min}(E)={\rm min}[\,     s_{14}(E),\,      
\theta(\varepsilon_2-E)s_{12}(E)  \nonumber\\
+\theta(E-\varepsilon_2)\theta(\varepsilon_3-E)s_{23}(E)+\theta(E-\varepsilon_3)s_{34}(E),
\nonumber\\
\theta(\varepsilon_2-E)s_{12}(E) +\theta(E-\varepsilon_2)s_{24}(E),  \nonumber\\
\theta(\varepsilon_3-E)s_{13}(E) +\theta(E-\varepsilon_3)s_{34}(E)  \,]
\label{werwolf1}
\end{gather}
where $\varepsilon_1\leq E\leq \varepsilon_4$.
Generalizations to $n>4$ are guessed from (\ref{werwolf}, \ref{werwolf1}). 

\begin{table*}[ht]
\begin{tabular}{|c|c|}
  \hline
Bargaining theory    & Statistical mechanics    \\
  \hline  \hline
Utilities of players & Entropy and minus energy \\
\hline
Joint actions of players & Probabilities of states for the physical system \\
\hline
Feasible set of utility values & Entropy-energy diagram \\
\hline
Defection point & Initial state \\
\hline
Pareto set & Maximum entropy curve for \\
            & positive inverse temperatures $\beta>0$ \\
\hline
\end{tabular}
\caption{ {\bf Bargaining theory $-$ statistical mechanics dictionary.} \\
Utilities (payoffs) are normally dimensionless and are defined
subjectively, via preferences of a given agent (player) \cite{luce-1,rubin,roth_book-1,karlin-1}. Entropy and
energy are physical, dimensional quantities \cite{grandy-1,balian-1}. The player's utility
$v_k(x_1,x_2)$ ($k=I, II$ denote the first and second player,
respectively) depends on the actions $x_1$ and $x_2$ taken by the first
and second player respectively \cite{luce-1,rubin,roth_book-1}.  Entropy and energy depend on the
probability of various states of the physical system \cite{grandy-1,balian-1}. Unlike actions,
these probabilities do not naturally fraction into two different
components. Hence it is unclear how to apply the non-cooperative game
theory to statistical mechanics. For cooperative game theory this problem is absent, since the actions are not separated.\\ 
All utility values from the feasible
set are potential outcomes of bargaining.  The feasible set is normally
convex and it is even modified to be comprehensive \cite{luce-1,rubin,roth_book-1}, i.e.  if
$(v_I,v_{II})$ belongs to the feasible set, then all points
$(w_I,w_{II})$ with $w_I\leq v_I$ and $w_{II}\leq v_{II}$ also belong to
the feasible set. The features of comprehensivity is motivated by the
observation that (arbitrary) worse utility values can be added to the
existing feasible set \cite{rubin,roth_book-1}.  The observation is not at all obvious (or
innocent) even in the game-theoretic context \cite{luce-1}. \\ The entropy-energy
diagram is a well-defined physical object that does not allow any (more
or less arbitrary) modification. It is not convex due to the minimal
entropy curve; see Fig.~\ref{fig6}. \\ Game-theoretically, the defection point is a specific
value of utilities which the players get if they fail to reach any
cooperation and/or agreement \cite{luce-1,rubin,roth_book-1}.  Normally, the defection point corresponds
to guaranteed payoffs of the players. The defection point does not have
any direct physical analogy. Instead of it we need to employ the notion
of the initial point that as such does not have game-theoretic
analogies (at least within axiomatic bargaining).\\ For a give set of utilities $(v_I,v_{II})\in {\cal F}$
belonging to the feasiblity set ${\cal F}$, the Pareto set ${\cal P}$
is defined as a subset of ${\cal F}$ such that $(\hat v_I,\hat
v_{II})\in {\cal P}$ if there does not exist any $(v_I,v_{II})\in {\cal
F}$ with $v_I\geq \hat v_I$ and $v_{II}\geq \hat v_{II}$, where at least
one of inequalities is strict \cite{luce-1,rubin,roth_book-1}. Thus there are no utility values 
that are jointly better than any point from the Pareto set. \\
If $(x_1,x_2)$ vary over a convex set, and if both
$v_I(x_1,x_2)$ and $v_{II}(x_1,x_2)$ are concave functions of $(x_1,x_2)$, 
then every point from ${\cal P}$ can be recovered from conditional maximization of $v_I(x_1,x_2)$
with a fixed $v_{II}(x_1,x_2)$ or {\it vice versa} \cite{karlin-1} (Karlin's lemma). \\
Within statistical physics the Pareto line coincides with the branch of the maximum entropy curve with a positive inverse temperatures $\beta>0$, because this branch
is also the minimum energy curve for a fixed entropy. Both entropy and energy are concave functions of probability, and the probability itself is defined over 
a convex set (simplex). Hence Karlin's lemma applies. \\
By analogy to the Pareto set we can define also 
the anti-Pareto set, which (as the opposite to the Pareto set) relates to ``worst" points of ${\cal F}$.  
The anti-Pareto set $\bar{{\cal P}}$
is defined as a subset of ${\cal F}$ such that $(\bar v_I,\bar
v_{II})\in \bar{{\cal P}}$ if there does not exist any $(v_I,v_{II})\in {\cal
F}$ with $v_I\leq \bar v_I$ and $v_{II}\leq \bar v_{II}$, where at least
one of inequalities is strict. On the entropy-energy diagram the only anti-Pareto point coincides with $(U={\rm min}[\,\{-\varepsilon_k\}_{k=1}^n\,],\, S=0)$.
After imposing Axiom 1 and going to the new set of feasible states, the unique anti-Pareto point coincides with the initial state. 
} 
\label{tab1}
\end{table*}

\subsection{2. Features of the Nash solution (\ref{bolo})}

\subsubsection{2.1 Concavity}

Let us write (\ref{bolo}) in coordinates (\ref{ojgur}):
\BEA
u_N={\rm argmax}_u[\, us(u)  \,],
\EEA
where (for obvious reasons) the maximization was already restricted
to the maximum entropy curve $s(u)$. Now recall that $s(u)$ is a
concave function. Local maxima of $us(u)$ are found from
[$\frac{\d s}{\d u}\equiv s'(u)$]:
\BEA
\label{oro}
[\, us(u)  \,]'|_{u=u_N}=u_Ns'(u_N)+s(u_N)=0. 
\EEA
Calculating 
\BEA
[\, us(u)  \,]''|_{u=u_N} =
-\frac{2s(u_N)}{u_N}+us''(u_N)<0
\EEA
we see that solutions $u_N$ of (\ref{oro}) are indeed local 
maxima due to $s''(u)<0$ (concavity) and $u\geq 0$, $s(u)\geq 0$, as seen from (\ref{ojgur}).

We shall now show that this local maximum is unique and hence coincides with the global maximum.
For any concave function $s(u)$ we have for $u_1\not=u_2$
\BEA
s(u_1)-s(u_2)<s'(u_2)(u_1-u_2),
\EEA
which produces after re-working and using (\ref{oro}) for $u\not=u_N$:
\BEA
us(u)-u_Ns(u_N)<s'(u_N)(u-u_N)^2\nonumber\\
=-\frac{s(u_N)}{u_N}\,(u-u_N)^2<0.
\EEA
Hence $u_N$ is the unique global maximum of $us(u)$.

\subsubsection{2.2 Comments on the textbook derivation of (\ref{bolo}).} 

\begin{figure}[ht]
\includegraphics[width=6cm]{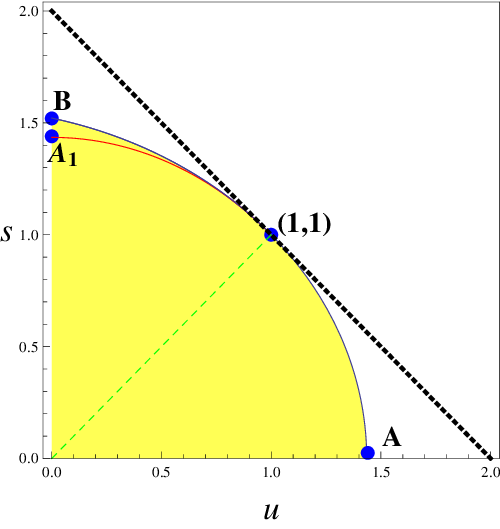}
\caption{Entropy-energy diagram in coordinates 
(\ref{ojgur}). Blue curve: $s(u)$. The affine freedom is chosen such that
the Nash solution (\ref{bolo})
coincides with the point $(1,1)$. The original domain of allowed states 
is filled in yellow. This domain is not symmetric with respect to $s=u$.
This is seen by looking at the inverse function $s^{-1}(u)$ [red line] of $s(u)$.
The domain below ${\bf AA_1}$ is symmetric with respect to $s=u$.
}
\label{fig4}
\end{figure} 

In the main text we emphasized that the derivation of the solution
(\ref{bolo}) for the axiomatic bargaining problem that was proposed by
Nash \cite{nash-1} and is reproduced in textbooks
\cite{luce-1,rubin,roth_book-1} has a serious deficiency. Namely,
(\ref{bolo}) is derived under an additional assumption, {\it viz.} that
one can enlarge the domain of allowed state on which the solution is
searched for. This is a drawback already in the game-theoretic
set-up, because it means that the payoffs of the original game are
(arbitrarily) modified. In contrast, restricting the domain of available
states can be motivated by forbiding certain probabilistic states 
(i.e. joint actions of the original game), which
can and should be viewed as a possible part of negotiations into which
the players engage.  For physical applications this assumption is
especially unwarranted, since it means that the original (physical)
entropy-energy phase diagram is arbitrarily modified. 

We now demonstrate on the example of Fig.~\ref{fig4} how specifically
this assumption is implemented. Fig.~\ref{fig4} shows an entropy-energy
diagram in relative coordinates (\ref{ojgur}) with $s(u)$ being the
maximum entropy curve.  The affine transformation were chosen such that
the Nash solution (\ref{bolo}) coincides with the point $(1,1)$. Now
recall (\ref{oro}).  Once $s(u_N)=u_N=1$, then $\frac{\d s(u_N)}{\d
u}=-1$, and since $s(u)$ is a concave function, then all allowed states
lay below the line $2-u$; see Fig.~\ref{fig4}.  If now one considers a
domain of {\it all} states $(u\geq 0,s\geq 0)$ (excluding the initial
point $(0,0)$) lying below the line $2-u$ (this is {\it the} problematic
move!), then $(1,1)$ is the unique solution in that larger domain.
Moving back to the original domain and applying the contraction
invariance axiom, we get that $(1,1)$ is the solution of the original
problem. 


\subsection{3. Maximum entropy method and its open problem}

\subsubsection{3.1 Maximum entropy method: a reminder}

The method originated in the cross-link between information theory and
statistical mechanics \cite{jaynes-1}. It applies well to
quasi-equilibrium statistical mechanics \cite{grandy-1,balian-1}, and
developed to become an inference method (for recovering unknown
probabilities) with a wide range of applications; see e.g.
\cite{grandy-1,balian-1,maxent_action-1,sbornik1}. 

As a brief reminder: let we do not know probabilities (\ref{simplex}), but
we happen to known that they hold a constraint:
\BEA
\label{uo}
U=U[p]\equiv {\sum}_{k=1}^nu_kp_k,
\EEA
with $\{u_k\}_{k=1}^n$ being realizations of some random 
variable ${\cal U}$. We refrain from calling ${\cal U}$ energy (or minus energy), since
applications of the method are general. 

Now if we know {\it precisely} the average $U$ in (\ref{uo}), then
unknown probabilities (\ref{simplex}) can be recovered from maximizing
the entropy (\ref{entro}) under constraint (\ref{uo}).  In a
well-defined sense this amounts to minimizing the number of assumption
to be made additionally for recovering probabilities
\cite{maxent_action-1}. The outcome of the maximization is well-known
and was already given by us in the main text:
\BEA
\label{gi-1}
\pi_i=e^{-\beta u_i}/Z, \quad Z={\sum}_{i=1}^n e^{-\beta u_i},
\EEA
where the Lagrange multiplier $\beta$ is determined from (\ref{uo}). 

The method has a number of desirable features \cite{maxent_action-1}. It
also has several derivations reviewed in
\cite{grandy-1,balian-1,maxent_action-1}. Importantly, the method is
independent from other inference practices, though it does have
relations with Bayesian statistics \cite{skyrms-1,enk-1,cheeseman_1-1}
and causal decision making \cite{skyrms-1}. 

\subsubsection{3.2 The open problem}

But from where we could know $U$ in (\ref{uo}) precisely? This can
happen in those (relatively rare) cases when our knowledge is based on
some symmetry or a law of nature. Otherwise, we have to know $U$ from a
finite number of experiments or|within subjective probability and
management science \cite{buckley-1}|from an expert opinion. The former
method will never provide us with a precise value of $U$, simply because
the number of experiments is finite. Opinions coming from experts do
naturally have certain uncertainty, or there can be at least two
slightly different expert opinions that are relevant for the
decision maker. Thus in all those case we can stick to a weaker form of
prior information, {\it viz.} that $U$ is known to belong to a certain
interval 
\BEA
\label{bars}
U\in [U_1,U_2], \qquad U_1<U_2. 
\EEA

This problem was recognized by the founder of the method
\cite{jaynes-2}, who did not offer any specific solution for it. Further studies 
attempted to solve the problem in several different ways:

-- Following Ref.~\cite{thomas-1}, which studies the entropy
maximization under more general type of constraints (not just a fixed
average), one can first fix $U$ by (\ref{uo}), calculate the maximum entropy $S(U)$,
and then maximize $S(U)$ over $U\in [U_1,U_2]$, which will mean 
maximizing entropy (\ref{entro}) under constraint (\ref{bars}). 
This produces:
\BEA
\label{batu}
{\rm max}_{p,\, U\in [U_1,U_2]} S[p] &=&S(U_1)\quad {\rm for} \quad U_1\geq U_{\rm av}\\
                                            &=&S(U_2)\quad {\rm for} \quad U_2\leq U_{\rm av} \\
                                            &=& \ln n\quad {\rm for} \quad U_1<U_{\rm av}<U_2,~~ \\
U_{\rm av}&=&\frac{1}{n}{\sum}_{k=1}^nu_k.
\label{ast}
\EEA
Such a solution is not acceptable; e.g. in the regime (\ref{batu}) it does not change when increasing $U_2$.
I.e. the solution does not feel the actual range of uncertainty implied in (\ref{bars}).

-- What is wrong with the simplest possibility that will state
$S(\frac{U_1+U_2}{2})$|i.e. the maximum entropy at the center of the
interval|as the solution to the problem? Taking the arithmetic average
of the interval independently from the underlying problem seems
arbitrary. 

Another issue is that each value of $U$ from the interval $[U_1,U_2]$ is
mapped to the (maximum) entropy value making up an interval of entropy
values. For $U_1\geq U_{\rm av}$ this interval is $[S_2,S_1]$, where
$S_1=S(U_1)$ and $S_2=S(U_2)$. Denoting by $U(S)$ the inverse function
of $S(U)$, we can take $U(\frac{\hat S_1+\hat S_2}{2})$ instead of
$S(\frac{U_1+U_2}{2})$. 

-- Ref.~\cite{cheeseman_2-1} assumes that (though the precise value of
the average is not known) we have a probability density $\rho(U)$ for
$U$. Following obvious rules, the knowledge of $\rho(U)$ translates into
the joint density $\rho(\pi_1,...,\pi_n)$ for maximum-entropy
probabilities (\ref{gi-1}). While this is technically well-defined, it is
not completely clear what is the meaning of probability density
$\rho(U)$ over the average $U$. One possibility is that the random
variable ${\cal U}$ is sampled independently $M$ times ($M$ is
necessarily finite), and the probability density of the empiric mean
\BEA
\label{barat}
\frac{1}{M}\sum_{\alpha=1}^M u_{[\alpha]},
\EEA
is identified with $\rho(U)$. This possibility is however problematic,
since it directly relates probability of the empiric mean with the
probability of the average. E.g. if we shall sample independently $M$
times from (\ref{gi-1}), then the average of the empiric mean equals
$\int\d U\,U\rho(U)$, but the empiric mean itself is not distributed via
$\rho(U)$.

-- Given (\ref{gi-1}), one can regard $\beta$ as an unknown parameter,
and then apply standard statistical methods for estimating it
\cite{rau-1}.  Thus within this solution the maximum entropy method is
not generalized: its standard outcome serves as the initial point for
applying standard tools of statistics. This is against the spirit of the
maximum entropy method that is meant to be an independent inference
principle \cite{jaynes-2}. 

-- Yet another route for solving the problem was discussed in
Ref.~\cite{jaynes-2}. (We mention this possibility, also because it came
out as the first reaction when discussing the above open problem with
practioners of the maximum entropy method \cite{roger}.) It amounts to
the situation, where in addition to the empiric mean (\ref{barat}) one
also fixes the second empiric moment $\frac{1}{M}\sum_{\alpha=1}^M
u_{[\alpha]}^{2}$ as the second constraint in maximizing (\ref{entro}).
It is hoped that since identifying the sample mean (\ref{barat}) with
the average $U$ is not sufficiently precise for a finite $M$, then
fixing the second moment will account for this lack of precision.  This
suggestion was not worked out in detail, but it is clear that it cannot
be relevant to the question we are interested in. Indeed, its
implementation will amount to fixing two different constraints, i.e. in
addition to knowing precisely the average $U$ in (\ref{uo}), it will
also fix the second moment $U'={\sum}_{k=1}^nu_k^2p_k$ thereby assuming
more information than the precise knowledge of $U$ entails.

\begin{figure}[ht]
\includegraphics[width=6cm]{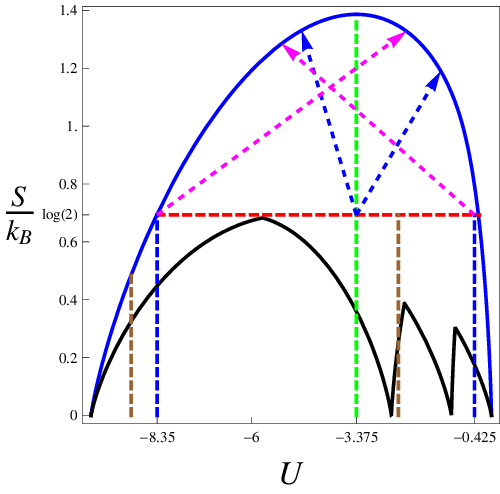}
\caption{Entropy-energy diagram; cf.~Fig.~\ref{fig1}. 
Entropy is $S$ and the minus energy $U=-E$ for 4-level system with energies [cf. the discussion around (\ref{simpo})]:
$\varepsilon_1=0$, $\varepsilon_2=1$, $\varepsilon_3=2.5$, and
$\varepsilon_4=10$. Maximal (minimal) entropy curves are denoted by blue
(black). All physically acceptable values of entropy and energy are
inside of the domain bounded by blue and black curves. Red dashed line denotes $\ln 2$.
Green dashed curve shows $U_{\rm av}$; see (\ref{gop}).\\
This figure illustrates the generalized maximum entropy method discussed in \S 3 and \S 4. 
Brown dashed lines indicate on specific values for $U_1$ and $U_2$ ($U_2<U_1$) which are not allowed according to axiom 2.\\
The right [left] blue dashed arrows points from the initial state $(U_{\rm i},S_{\rm i})=(-3.375,\ln 2)$ given by 
(\ref{ko-2}) [by (\ref{ko-3})] to the corresponding Nash solutions (\ref{koko2}) [(\ref{gaga2})] defined over domain 
$(U\geq U_{\rm i}, S\geq S_{\rm i})$ [over $(U\leq U_{\rm i}, S\geq S_{\rm i})$]. \\
The interval $(U_1,U_2)=(-8.35,-0.425)$ is not allowed, because there are two possibilities for initial points 
$(U_{\rm i},S_{\rm i})=(-8.35,\ln 2)$ and $(U_{\rm i},S_{\rm i})=(-0.425,\ln 2)$, each one producing its own Nash solution 
shown by magenta dashed arrows; cf.~discussion around (\ref{orujan}).
}
\label{fig6}
\end{figure} 

\subsubsection{3.3 Solving the problem via bargaining}

Main premises of this solution is that we should simultaneously account
for both the uncertainty in $U$ and in $S$, and that we should account
for the duality of optimization, i.e. that the maximum
entropy result can be also obtained via the optimization of $U[p]$ under
a fixed $S=S[p]$.  

Let us first of all add an additional restriction in
(\ref{bars}):
\begin{gather}
\label{bars2}
U\in [U_1,U_2], \qquad U_{\rm av}\equiv \frac{1}{n}{\sum}_{k=1}^nu_k <U_1<U_2,\\
S(U)\equiv {\rm max}_{p,\, U=U[p]} S[p].
\end{gather}
The case $U_{\rm av}>U_2$ is treated similarly to (\ref{bars2}) with obvious generalizations explained below. 
The general case (\ref{bars}) is more difficult and will be addressed at the end of the next chapter. 

We now know that the maximum entropy solution (\ref{gi-1}) can be recovered also by maximizing $U$ over $\{p_k\}_{k=1}^n$
for a fixed entropy. Note that the uncertainty (\ref{bars2}) translates into an uncertainty 
\BEA
\label{ben}
S\in [S_2,S_1], \qquad S_1\equiv S(U_1), \quad S_2\equiv S(U_2), 
\EEA
in the maximum entropy. We now take the joint uncertainty domain $\Omega$ in the $(U,S)$ diagram.
$\Omega$ includes all points
$(\bar{U},\bar{S})$, where $\bar{U}\in [U_1,U_2]$ [see (\ref{bars2})], and where $\bar{S}\in [S_2,S_1]$; see (\ref{ben}). 
$\Omega$ has the structure of the domain required by Axiom 1, with the (initial) point
\BEA
\label{ko-2}
(U_{\rm i}, S_{\rm i})= (U_1,S_2)
\EEA
being the unique anti-Pareto point, i.e. the unique 
point, where $U$ and $S$ jointly minimize. Recall the definition of the anti-Pareto
set given in the caption of Table I. 

We shall take Axiom 1 in a slightly restricted form 
as compared with its original form (Axiom 1) given by
(\ref{sumo}) of the main text
\footnote{The reason of this restriction is that for being able to
deduce thermalization using only the restricted affine-covariance
(\ref{a1-1}), we anyhow need to require the strict inequality $U_{\rm
i}< U_{\rm f}$; cf the discussion after (\ref{ojgur}). }
\BEA
{\it Axiom~ 1':}\quad U_{\rm i}< U_{\rm f}, \qquad S_{\rm i}< S_{\rm f}.
\label{sumo-2}
\EEA
Axioms 2 and 4 go on as stated in the main text. In particular, 
Axiom 2 ensures that $\Omega$ is a convex domain, i.e. that it does
not hit the minimum entropy curve. 

However, the application of axiom 3 on the affine covariance
needs a restriction, because it is seen from (\ref{ko-2}) that the
initial point $(U_1,S_2)=(U_1,S(U_2))$ does transform in the affine way
upon affine transformations $U \to  a^{-1} U +d$ of $U$; cf (\ref{a1}). 
On the other hand, is obviously intact under affine transformations of $S$
that we take as the restricted form of Axiom 3 \footnote{
We emphasize that for the present problem|where we have an uncertainty
interval $[U_1,U_2]$ for $U$|the inapplicability of the affine
transformation $U \to a^{-1} U +d$ is expected for at least two reasons.
Firstly, the uncertainty interval $[U_1,U_2]$ does generally change
under this transformation, which indicates on altogether a different
problem.  Secondly, the very definition of the initial point
(\ref{ko-2}) does already connect with the maximum entropy
curve; hence we do not expect the full freedom with respect to affine
transformations to retain. 
}:
\BEA
\label{a1-1}
{\it Axiom~ 3':}\quad S\to b^{-1} S +c ,\qquad b>0.
\EEA
It should be clear from our discussion in the main text|cf.~discussions
after (\ref{ojgur}) and after (\ref{bolo})|that Axioms 1', 2, 3' and 4
suffice for deriving thermalization. Note that the discussion after
(\ref{bolo}) of the main text assumed only affine transformations of $U$
only, but we could consider affine transformations of $S$ only with the
same success; see Fig.~\ref{fig5}. Note as well that instead of affine
transformation (\ref{a1-1}) we can apply $U \to a^{-1} U +d$ [i.e. we
can reformulate axiom (\ref{a1-1})], if the initial state (\ref{ko-2})
is parametrized as $(U_{\rm i}, S_{\rm i})= (U_1,S_2)=(U(S_1),S_2)$ via
the inverse function $U(S)={\rm max}_{p,\, S=S[p] }U[p]$ of $S(U)$. 

Axiom 5 will go as stated in the
main text and demands symmetry between maximizing $S$ and maximizing
$U$. Altogether, Axioms 1', 2, 3', 4 and 5 suffice for deriving
\BEA
\label{koko2}
{\rm argmax}_{(U,S)}\left[  (U-U_{1})(S-S_2) \right],
\EEA
as the solution of the problem with uncertainty interval (\ref{bars2}).

\subsubsection{3.4 Generalizing the solution to other types of uncertainty intervals}

Uncertainty intervals that instead of (\ref{bars2}) hold 
\BEA
\label{bars3}
U\in [U_1,U_2], \qquad U_1<U_2<U_{\rm av},
\EEA
are straightforward to deal with. Now relevant points on the maximum entropy curve can be reached
by entropy maximization for a fixed $U=U[p]$, or {\it minimizing} $U[p]$ for a fixed entropy $S=S[p]$. 
Hence the initial point and Axiom 1' now read [cf.~(\ref{ko-2}, \ref{sumo-2})]:
\BEA
\label{ko-3}
&& (U_{\rm i}, S_{\rm i})= (U_2,S_1),\\
&& {\it Axiom~ 1'':}\quad U_{\rm i}> U_{\rm f}, \qquad S_{\rm i}< S_{\rm f}.
\label{sumo-3}
\EEA
Instead of (\ref{koko2}), the solution under (\ref{bars3}, \ref{ko-3})
will read:
\BEA
{\rm argmax}_{(U,S)}\left[  (U_{2}-U)(S-S_1) \right].
\label{gaga2}
\EEA

Let us now take the case, where 
$[U_1,U_2]$ holds neither (\ref{bars2}), nor (\ref{bars3}), i.e. it holds:
\BEA
\label{bars4}
U\in [U_1,U_2], \qquad U_1<U_{\rm av}<U_2.
\EEA
Now we do not know a priori whether we should 
maximize or minimize over $U[p]$. 

We define $\Omega$ as above by joining together uncertainties of $U$ and
$E$, i.e. by including all points $(\bar{U},\bar{S})$, where $\bar{U}\in
[U_1,U_2]$ [see (\ref{bars4})], but where $\bar{S}\in \left(\, {\rm
min}[S_1,S_2],\,\ln n\right)$.  The latter interval, due to
(\ref{bars4}), is where the maximum entropy values are contained.  It is
clear that generically $\Omega$ has a structure of a right ``triangle"
formed by by two legs and a convex curve instead of the hypotenuse.
Depending on whether $S_2<S_1$ or $S_1<S_2$, we apply to $\Omega$ either
(\ref{ko-2}) with Axiom 1' (\ref{sumo-2}), or (\ref{ko-3}) with Axiom
1'' (\ref{sumo-3}). Axioms 2, 3', 4, 5 apply without changes.
Hence the initial point and solution reads, respectively in two regimes:
\begin{align}
 {\rm For} \quad S_2<S_1:\quad & (U_{\rm i}, S_{\rm i})= (U_1,S_2),\\
\label{amba1}
                         & {\rm argmax}_{(U,S)}\left[  (U-U_{1})(S-S_2) \right].\\
 {\rm For} \quad S_1<S_2: \quad & (U_{\rm i}, S_{\rm i})= (U_2,S_1),\\
\label{amba2}
                         & {\rm argmax}_{(U,S)}\left[  (U_{2}-U)(S-S_1) \right].
\end{align}
There is only one case, where solutions (\ref{amba1}, \ref{amba2}) become umbiguous:
\BEA
\label{orujan}
S_2=S( U_2>0)=S_1=S( U_1<0).
\EEA
Indeed, now interpolating from $S_2<S_1$ will lead to (\ref{amba1}), which is different
as compared with interpolating (\ref{amba2}) from $S_1<S_2$.
Thus we state that under (\ref{orujan}) the prior information (\ref{bars4}) 
does not suffice for drawing a unique conclusion with the bargaining method. 

Fig.~\ref{fig6} illustrates all the above solution on  
an entropy-energy diagram. Two examples of the set
$\Omega$ are shown by blue, green (dashed) and red (dashed) lines.
Brown dashed lines show an example of $(U_1,U_2)$ ($U_1<U_2$), where the
domain of states allowed according to Axiom 1 is not convex: it will
cross the minimum entropy curve $S_{\rm min}(U)$ denoted by black lines
on Fig.~\ref{fig6}. Hence such values of $(U_1,U_2)$ are not allowed.
Fig.~\ref{fig6} shows two examples of allowed intervals $(U_1,U_2)$:
$(U_1,U_2)=(-8.35, -3.375)$ and $(U_1,U_2)=(-3.375, -0.425)$. The
corresponding values of $S_1=S(U_1)$ and $S_2=S(U_2)$ are $(\ln 2, \ln
4)$ and $(\ln 4, \ln 2)$. Blue arrows join the initial states with
corresponding Nash solutions (\ref{koko2}, \ref{gaga2}). 

It is seen that for $(U_1,U_2)=(-8.35, -0.425)$, where $S_1=S_2=\ln 2$,
there are two Nash solutions denoted by dashed magenta arrows on
Fig.~\ref{fig6}. For this case (\ref{orujan}) the existing prior
information does not allow to single out a unique solution.


\begin{thebibliography}{99}
\bibitem{foo1} While our work is among first applications of game theory
to statistical physics, connections between mathematical economics, thermodynamics and
econophysics are well-known
\cite{saslow,candeal,tsirlin,smith,yakovenko,bruce}; e.g. there are
interesting analogies between the axiomatic features of entropy and
utility \cite{candeal}. For application of statistical physics idea in evolutionary game theory see 
\cite{chatelier}.
\bibitem{foo2} Axiomatic approaches to equilibrium thermodynamics have a
long history; see \cite{callen,buchdahl,cooper,lieb}. They 
revolve around axiomatic introduction of entropy, whereas in our
situation the entropy is just assumed to hold its standard form.
\end{thebibliography}
\end{document}